\documentclass[
nofootinbib,
 amsmath,amssymb,
 aps,
prl,
twocolumn, superscriptaddress
]{revtex4-1}

\usepackage[normalem]{ulem}
\usepackage{graphicx}
\usepackage{dcolumn}
\usepackage{bm}
\usepackage{hyperref}

\usepackage{xcolor}
\usepackage{stackengine}

\begin{document}

\title{Parametric Resonance in the Einstein frame: the Jordan-frame Doppelgänger}
\author{Karim H. Seleim}
\email{khammam@zewailcity.edu.eg}
\affiliation{University of Science and Technology, Zewail City of Science and Technology, 6th of October City, Giza 12588, Egypt}
\author{Richa Arya}
\email{richaarya@mnnit.ac.in}
\affiliation{Department of Physics, Motilal Nehru National Institute of Technology Allahabad, Prayagraj 211004, U.P.,  India}
\author{Sergio E. Jor\'as}
\email{joras@if.ufrj.br}
\affiliation{
 Instituto de F\'\i sica, Universidade Federal do Rio de Janeiro,\\
 CEP 21941-972 Rio de Janeiro, RJ, Brazil}
                                        
\begin{abstract}
Modified $f(R)$ theories of gravity have been investigated for quite a long time in the literature as a possible explanation for the inflationary period of the universe. The correspondence to General Relativity with an extra scalar field $\tilde\phi$ in the so-called Einstein Frame via a conformal transformation is a major tool in this class of theories. Here, we assume three different potentials $V(\tilde\phi)$ and a parametric-resonance coupling between $\tilde\phi$ and a secondary scalar field $\tilde\psi$ such that one can have both inflation and preheating in the Einstein frame. We study the instability resonance band structure for our models. Further, we determine the correspondent mechanism --- and the function $f(R)$ itself --- in the Jordan frame, that is possibly related to the so-called vacuum awakening mechanism. 
\end{abstract}

\maketitle

Modified theories of gravity have been studied in different energy scales, with a common goal: 
accelerating the expansion of the universe. In the early universe, modified equations do not require an extra (yet unknown) inflaton field $\phi$. In the current universe, they aim to replace the cosmological constant $\Lambda$ \cite{Sotiriou:2006hs,Sotiriou:2008rp,DeFelice:2010aj,Joras:2011zz}.

Most of the simplest extensions (in particular, polynomial expressions in the Ricci scalar $R$) have already been discarded by constraints from the Solar system, the background cosmological evolution, the growth rate of cosmological perturbations, and the stability of relativistic stars. Note, however, that some modifications do work quite well in limited scenarios, such as Starobinsky model $f(R) = R + \alpha R^2$ \cite{Starobinsky} --- which is a strong candidate for generating primordial inflation \cite{Planck}, but it is also strongly disfavored for standard cosmology \cite{luca} and stellar stability \cite{Pretel:2020rqx}. 

Here, we focus on the cosmological evolution of the primordial universe and, in particular, on the preheating phase that takes place right after the end of supercooled inflation. As such, it is responsible for (pre)heating the universe to temperatures compatible with the standard Big Bang scenario: namely, a hot dense radiation-dominated background. Instead of following a systematic search through an infinite-dimensional function space in a quest for the ``correct" $f(R)$, here we take an alternative approach:  we start in the Einstein Frame (EF) \cite{Faraoni:1999hp,Quiros:2012rnn,Postma:2014vaa,Racioppi:2021jai}, where the extra degree of freedom is materialized in a scalar field $\tilde\phi$ subject to a potential $V(\tilde\phi)$ and look for the corresponding $f(R)$ in the Jordan Frame (JF). We are particularly interested in the interaction between a secondary scalar field $\tilde\psi$ (also defined in the EF) and $\tilde\phi$, and their corresponding interaction term in the JF.

The question we seek to investigate in the present work is threefold: First, is there a preheating mechanism \cite{Greene:1997fu,Bassett:1997az,Tsujikawa:1999me,Tsujikawa:1999jh,Bertolami:2010ke,Tsujikawa:1999iv,Watanabe:2006ku}, i.e, one that relies on parametric resonance (PR), between the secondary field $\tilde \psi$ and the ``inflaton'' $\tilde\phi$ in the EF? Secondly, if so, what is the corresponding mechanism in the JF that would then explain the exponential growth of a given set of modes of the field $\psi$? Thirdly, what is the corresponding $f(R)$ and does it follow the usual constraints (such as stability, etc.)?

The present work differs from previous ones \cite{1999PhRvD..60l3505T, Tsujikawa:1999iv, 2017IJMPD..2650152V, Abramo_2002,2023ForPh..7100041K} in two fundamental aspects: First, our conformal transformation does not depend on the matter field $\psi$ (as opposed to Ref.~\cite{1999PhRvD..60l3505T}). Therefore, here the $\tilde\phi$ field still formally belongs to the gravity sector of the theory. Secondly, we assume only an approximately conformal-like term in the JF, as opposed to Ref.~\cite{2017IJMPD..2650152V}, where a parametric-resonance-like term is assumed from the beginning. i.e, in the JF. Besides, the discussion in the EF  we present here is deeper than the one presented in the latter reference; here, we are able to present a full correspondence between the amplification mechanisms in each frame. In Ref.~\cite{Abramo_2002}, the authors also study conformal-coupling terms in the JF, but do not investigate the parametric resonance, which is fully investigated in the present work. The authors of Ref.~\cite{2023ForPh..7100041K} do study parametric resonance in scalar-field theories of gravity, but we define the potential in the EF (as opposed to their definition, made in the JF) and, consequently, they do not take the final step to an explicit $f(R)$ function, as we do here. We believe that is an essential new ingredient to guide us towards the ``final" expression in the JF.

In the next sections we will briefly review the main topics of the present work: the Jordan and Einstein Frames, followed by the preheating mechanism (and its instability bands). Finally, we check the viability of the model.

\section{Jordan and Einstein Frames}

In the Jordan Frame (JF), where the energy-momentum tensor of any matter/radiation is conserved by itself,  one writes the Einstein-Hilbert Lagrangian as
\begin{equation}
{\cal L}_{GR} \equiv \sqrt{-g} \frac{1}{2\kappa^2}R,
\end{equation}
where $g\equiv \det(g_{\mu\nu})$, $R$ is the corresponding Ricci scalar and $\kappa^2 \equiv 8 \pi G_N \equiv M_{pl}^{-2}$. Still in the JF, one writes the modified theory of gravity as 
\begin{equation}
{\cal L}_f \equiv \sqrt{-g} \frac{1}{2\kappa^2} f(R) ,
\label{jordan}
\end{equation}
where $f(R)$ is a non-linear function of $R$ (otherwise, we would have GR again).  Upon a Legendre transformation --- where we replace $R$ by the dimensionless field $\phi(R)\equiv f'(R)\equiv df/dR$ as the independent variable ---  Eq.~(\ref{jordan}) is written 
\begin{equation}
{\cal L}_J = \sqrt{-g} \bigg[ \frac{1}{2\kappa^2} \phi R(\phi) - W(\phi)\bigg],
\label{legendre}
\end{equation}
where $ W(\phi) \equiv \{\phi R(\phi) - f[R(\phi)]\}/(2\kappa^2)$ and $R(\phi)$ is obtained from inverting the very definition of $\phi$. For that, we require that $f''(R) \equiv d^2f/dR^2 \neq 0$, except, perhaps, only at particular values. Note that Eq.~(\ref{legendre}) is {\bf not} a Brans-Dicke Lagrangian and that the field $\phi$, in spite of no kinetic term, does have an equation of motion of its own. This is a consequence of the extra degree of freedom inherent to $f(R)$ theories, as we recall below.

Upon a conformal transformation, defined by
\begin{equation}
g_{\mu\nu} \rightarrow \tilde{g}_{\mu\nu} \equiv \phi \cdot g_{\mu\nu},
\label{conformal}
\end{equation}
one arrives at the Einstein Frame (EF)\footnote{From now on, all the quantities defined in the EF are indicated by a ``$\sim$"  in superscript.}, where Eq.~(\ref{jordan}) is written (up to surface terms) as 
\begin{equation}
\tilde {\cal L} = \sqrt{-\tilde g} \bigg( \frac{1}{2\kappa^2}\tilde R - \frac{1}{2} \tilde g^{\alpha\beta} \partial_\alpha \tilde\phi \partial_\beta \tilde\phi -V(\tilde\phi)\bigg),
\label{LEphi}
\end{equation}
where $\tilde R $ is the Ricci scalar calculated from the metric $\tilde g _{\mu\nu}$ (whose determinant is $\tilde g$),   $\beta \tilde\phi \equiv \ln\phi$, $\beta\equiv \sqrt{2\kappa^2/3}$ and
\begin{equation}
V(\tilde\phi) \equiv \frac{R(\phi) \phi - f[R(\phi)]}{2\kappa^2\phi^2}.
\label{Vtilde}
\end{equation}
Of course, in the above equation, each and every $\phi$ has to be written in terms of $\tilde\phi$, namely $\phi=\exp[+\beta\tilde\phi]$.

Now we shall add to the JF Lagrangian (\ref{jordan})  extra terms corresponding to an extra massive scalar field $\psi$, minimally coupled to gravity:
\begin{align}
{\cal L}_\psi &\equiv   \sqrt{-g} \bigg(-\frac{1}{2} g^{\alpha\beta} \partial_\alpha \psi \partial_\beta \psi - V_\psi(\psi) \bigg)
\label{Lpsi}
\\
V_\psi(\psi) &\equiv \frac{1}{2} m_\psi^2 \psi^2.
\label{Vpsi}
\end{align}
Upon the same conformal transformation (\ref{conformal}), ${\cal L}_\psi$ becomes
\begin{align}
\tilde{\cal L}_\psi &=  \phi^{-2} \sqrt{-\tilde g}\,\bigg(-\frac{1}{2} \phi^{+1}\tilde g^{\alpha\beta} \partial_\alpha \psi \partial_\beta \psi - V_\psi(\psi) \bigg)\\
&=  \sqrt{-\tilde g}\bigg[-\frac{1}{2} \tilde g^{\alpha\beta} \phi^{-1} \,  \partial_\alpha \psi \partial_\beta \psi - \phi^{-2} V_\psi(\psi) \bigg]\\
&= \sqrt{-\tilde g} \bigg(-\frac{1}{2}\tilde g^{\alpha\beta} D_\alpha \tilde\psi D_\beta \tilde\psi - \frac{1}{2} \tilde m_\psi^2 {\tilde\psi}^2 \bigg),
\label{LEpsi}
\end{align}
where we have defined new tilde ($\tilde ~$) quantities and a new covariant derivative: 
\begin{align}
\tilde\psi &\equiv \phi^{-1/2} \psi \\
\tilde m_\psi^2 &\equiv \phi^{-1}\, m_\psi^2 \\
D_\alpha &\equiv \partial_\alpha + \frac{1}{2}\partial_\alpha(\beta \tilde\phi).
\end{align}
Note that this is similar to the covariant derivative of classical Electromagnetism (EM), but not quite the same. The reason is twofold: here, the (would-be EM) field $A_\mu\equiv \partial_\mu(\beta \tilde\phi)$ is Real (i.e, not Complex). Secondly, being a gradient, $A_\mu$ is pure gauge and, therefore, the corresponding  kinetic term $F_{\mu\nu}\equiv \partial_\mu A_\nu - \partial_\nu A_\mu $ would identically vanish.
It is worth noticing that, since $\phi \equiv f'>0$, the new field $\tilde \psi$ and its mass are well defined.

\section{Parametric Resonance}

Parametric Resonance (PR) is an ubiquitous phenomenon, showing up in a huge range of scales: from the localization of waves in almost-periodic crystals (the so-called Anderson Localization \cite{brandenberger2012towards}) 
up to the preheating process in the primordial universe, just after inflation. The common ingredient is an oscillating mass term, which can be achieved by an interacting term in the Lagrangian such as 
\begin{equation}
{\cal L}_{\rm int} = \frac{1}{2} \tilde \xi
\tilde\phi^n \tilde\psi^2,
\label{Lint}
\end{equation}
where $\tilde \xi$ is the coupling constant between fields in the EF (to be related to its counterpart in the JF in the next section)
so that the equation of motion for the modes $\tilde\psi_k$ is 
\begin{equation}
\ddot{\tilde \psi}_k + \left( k^2 + \tilde m_\psi^2 + \tilde \xi \tilde \phi^n \right) \tilde\psi_k = 0
\label{mathieu}
\end{equation}
where we have assumed a canonical kinetic term and a potential $V(\tilde\psi) = \frac{1}{2} \tilde m_{\psi}^2 \tilde\psi^2$ and neglected the expansion of the universe. If we further assume that $\tilde\phi$ is oscillating at the bottom of its own (approximately quadratic) potential, then the expression above is known as the Mathieu equation \cite{wolfram} for a function $x(z)$: 
\begin{equation}
\frac{d^2x}{dz^2} + [A - 2 q \cos(2z)] x(z) = 0,
\label{mathieu}
\end{equation}
where $z\equiv \tilde m_\phi t$, $A\equiv (k^2 + \tilde m_\psi^2)/\tilde m^2_\phi + 2 q$, and $q\equiv \tilde\xi \tilde\phi_o^2/(4 \tilde m^2_\phi)$ are constants and  $\phi_o$ is the initial amplitude of $\tilde\phi$. The corresponding solutions are exponentially increasing in time if the parameters $A$ and $q$ are inside one of the so-called instability bands in the parameter space $A \times q$ --- which, of course, depends on the wavenumber $k$, the field mass $\tilde m_\psi^2$ and the coupling $\tilde \xi$.

\section{Coupling fields in the EF} 

We would like to find out how such a coupling (\ref{Lint}) would be written in the JF. Obviously\footnote{Since $\tilde\phi$ corresponds to a non-trivial expression when written in terms of JF quantities.} it would generate a non-minimal coupling between $\psi$ and $R$-terms. The simplest one is a conformal-like- coupling term, namely
\begin{equation}
{\cal L}_{\rm int} = \xi R \psi^2,
\label{confcoup}
\end{equation}
which can be easily written in term of EF quantities, using the inverse mapping \cite{Magnano:1993bd}
\begin{align}
\label{fphi}
f(\tilde\phi) &= 3 \beta^2 {\rm e}^{2 \beta \tilde\phi} \left[V(\tilde\phi) + \frac{1}{\beta}  \frac{{\rm d} V(\tilde\phi)}{d\tilde\phi} \right] \quad {\rm and}\\
\label{Rphi}
R(\tilde\phi) &= 3 \beta^2 {\rm e}^{\beta \tilde\phi} \left[2 V(\tilde\phi) + \frac{1}{\beta} \frac{{\rm d} V(\tilde\phi)}{d\tilde\phi} \right].
\end{align}
Indeed, around the minimum of the potential at $\tilde\phi=0$, assuming a simple power-law for $V(\tilde\phi)\sim \tilde\phi^{2m}$ (with $m\geq 1$ for a stable minimum), Eq.~(\ref{Rphi}) yields
\begin{equation}
 R \psi^2 \sim  \tilde\phi^{2m-1}\psi^2
\end{equation}
as the first term in a series expansion. Except for extra global multiplicative factors from the conformal transformation, the right-hand-side of the expression above does correspond to a parametric-resonance (PR) term in the EF. Such coupling, along with an oscillatory $\tilde\phi$, is the basic ingredient for the preheating mechanism that follows inflation.

Let us then look for exact solutions, i.e, particular potentials $V(\tilde\phi$), for which 
\begin{align}
\tilde \xi \tilde\phi^n \tilde \psi^2 & = 
\xi R \psi^2 \\
& = 3 \xi \beta^2 {\rm e}^{\beta \tilde\phi} \left[2 V(\tilde\phi) + \frac{1}{\beta} \frac{{\rm d} V(\tilde\phi)}{d\tilde\phi} \right] \bigg(\tilde\psi^2 {\rm e}^{\beta \tilde\phi} \bigg) \nonumber\\
& = 3 \xi \beta^2 {\rm e}^{2 \beta \tilde\phi} \left[2 V(\tilde\phi) + \frac{1}{\beta} \frac{{\rm d} V(\tilde\phi)}{d\tilde\phi} \right] \tilde\psi^2
\label{map}
\end{align}
where $\tilde\xi = 3 \xi \beta^2$ and $n>0$ (to avoid singularities when $\tilde \phi =0$).
There is, indeed, an exact solution, given by the choice 
\begin{equation}
V_{\rm ex}(\tilde\phi)= {\rm e}^{ - 2 \beta  \tilde\phi}
\bigg[ C + \frac{1}{3(n+1)\beta^{n+2}} (\beta \tilde\phi)^{n+1}\bigg] ,
\label{Vex}
\end{equation}
where $C$ is an arbitrary constant. The corresponding potential, however,  does not feature a stable vacuum state (for $\tilde\phi<0$) for even $n$ --- see Fig.~{\ref{exact}). 
Besides, if we focus on the primordial universe --- in particular, the inflationary phase --- it does not yield a slow-roll phase either, even for odd $n$. 
Therefore, it does not yield a complete picture that could describe both inflation {\it and} preheating. 

\begin{figure}
\centering
\includegraphics[width=0.45\textwidth]{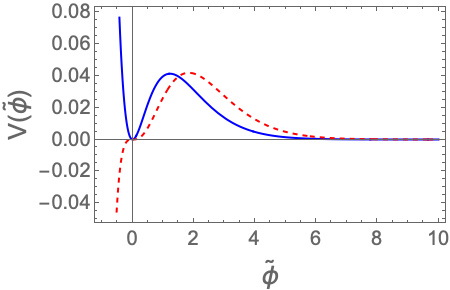}
\caption{Plot of $V_{\rm ex}(\tilde\phi)$, given by Eq.~(\ref{Vex}) for $C=0$ and $n=1$ (solid blue) and $n=2$ (dashed red). One can see that there is no flat plateau that would generate suitable slow-roll period for inflation and neither (for even $n$) a stable minimum around which the field would oscillate.}
\label{exact}
\end{figure}

Fortunately, we actually do not need an  exact solution, since the PR happens only at the bottom of the potential $V(\tilde\phi)$, not for large $\tilde\phi$ where inflation takes place.

Since we are looking at the full evolution of the $\tilde\phi$ field, we need to write down the full Lagrangian (for both $\tilde\phi$ and $\tilde\psi$). Upon a conformal transformation, a conformal-like term in the JF is written as in Eq.~(\ref{map}) in the EF, for a given potential $V(\tilde\phi)$. But what happens to the other terms in the Lagrangian, in the JF, upon the same conformal transformation? 
While the full Lagrangian in the JF is
\begin{equation}
{\cal L}_J =\sqrt{-g}\bigg[
f(R) - \frac{1}{2} g^{\alpha\beta} \partial_\alpha \psi \partial_\beta \psi - V_\psi(\psi) + \xi R \psi^2
\bigg],
\label{Lfull}
\end{equation}
the full Lagrangian in the EF, written from Eqs.~(\ref{LEphi}), (\ref{Vtilde}), (\ref{LEpsi}) and (\ref{map}), is
\begin{align}
{\cal L}_E= \sqrt{-\tilde g} &\bigg\{\frac{1}{2\kappa^2}\tilde R - \frac{1}{2}\tilde g^{\alpha\beta} \partial_\alpha \tilde\phi \partial_\beta \tilde\phi -V(\tilde\phi)+\nonumber\\
& - \frac{1}{2}\tilde g^{\alpha\beta} D_\alpha \tilde\psi D_\beta \tilde\psi - \frac{1}{2} \tilde m_\psi^2 {\tilde\psi}^2 + \nonumber \\
& + 3 \xi \beta^2 {\rm e}^{2 \beta \tilde\phi} \left[2 V(\tilde\phi) + \frac{1}{\beta} \frac{{\rm d} V(\tilde\phi)}{d\tilde\phi} \right] \tilde\psi^2 \bigg\}.
\end{align}
Any given $V(\tilde\phi$) is then assumed in the last line (which originated from the conformal-like term) and then approximated --- for $\beta\tilde\phi \ll 1$, i.e, around the bottom of $V(\tilde\phi)$, which then yields a conformal-like coupling between the fields $\tilde \phi$ and $\tilde\psi$ in the EF, as we will see below.
In that regime, as previously discussed, the expression above does behave as an oscillating effective mass for the $\tilde\psi$ field, which is the main characteristic for the PR.

To proceed, we assume three quite interesting cases that can be fully investigated analytically. 
The first two cases --- quadratic and quartic potentials --- are disfavoured by observational data from CMB \cite{Planck} as candidates to inflation. Nevertheless, they may be quite good approximations to more complex potentials close to the bottom at $\tilde\phi=0$, where PR takes place.

\subsection{Quadratic Potential}
The simplest potential with a minimum  is\footnote{This case and its Thermodynamics interpretation has been fully investigated previously \cite{Peralta:2019xlt}.} 
\begin{equation}
V_2(\tilde\phi) =\frac{1}{2} \tilde m_\phi^2 \tilde\phi^2,
\label{V2}
\end{equation}
which corresponds to a multi valued $f(R)$, plotted in Fig.~\ref{fRphi24e}, where we have used $\tilde m_\phi=1$. When the field is restricted to $\beta\tilde\phi>\beta\tilde\phi_*^{(2)}\equiv (\sqrt{5}-3)/2\approx -0.38$, only the quasi-linear branch is probed (the lower one, where $d^2f/dR^2<0$, is {\it not}). Within this range and close to the bottom of the potential, at $\tilde\phi=0$ (where the oscillating scalar field behaves as dust), one can expand Eq.~(\ref{Rphi}) in series, solve it for $R$ and use it in Eq.~(\ref{fphi}), which yields
\begin{equation}
f_2(R) \approx R + \frac{1}{6\tilde m_\phi^2} R^2 + \frac{1}{9\tilde m_\phi^4} R^3 + {\cal O}(R^4),
\label{f2}
\end{equation}
from which one can notice that this is {\it approximately} Starobinky's model only at very low $R$.  For the quadratic potential, the interaction term is then written
\begin{equation}
{\cal L}_{\rm int}^{(2)} = 3 \xi \tilde m_\phi^2 \beta\, \tilde\phi \tilde\psi^2 
= g_2 \, \tilde\phi \tilde\psi^2
\label{Lint2}
\end{equation}
where $g_2 \equiv 3 \xi \tilde m_\phi^2 \beta$.

\begin{figure}
\includegraphics[width=0.45\textwidth]{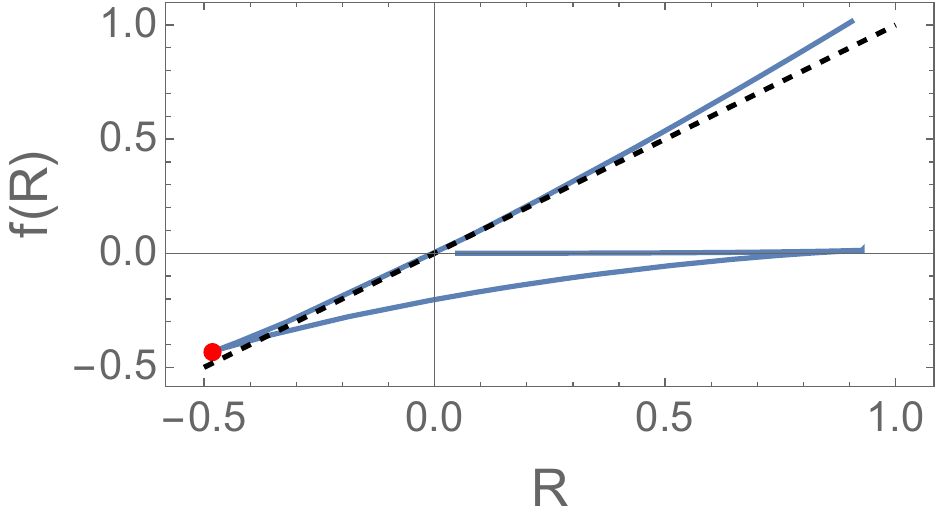}
\includegraphics[width=0.45\textwidth]{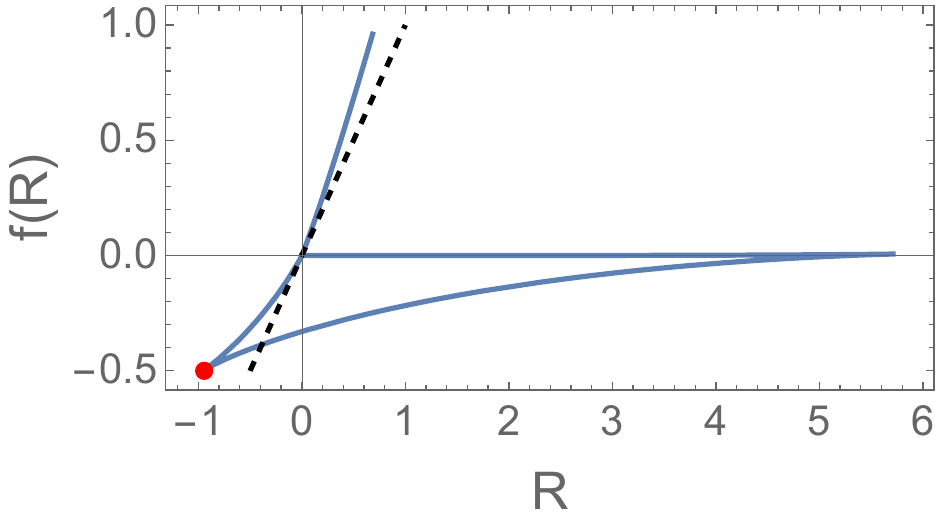}
\includegraphics[width=0.45\textwidth]{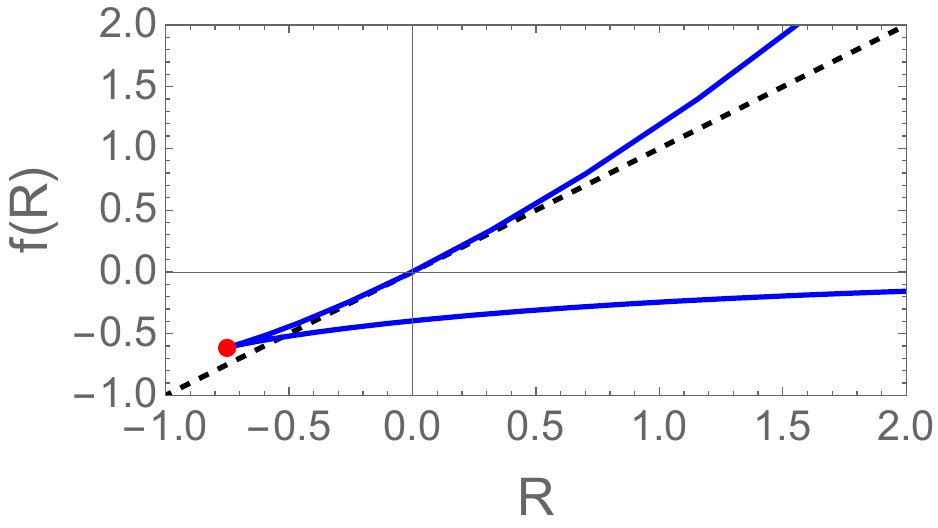}
\caption{Functions $f(R)$ given by: {\bf (upper panel)} corresponding to the quadratic potential, Eq.~(\ref{f2}),  {\bf (middle panel)} to the quartic potential, Eq.~(\ref{f4}), and {\bf (lower panel)} to the exponential one, Eq.~(\ref{fexp1}). In all panels, the dashed straight line is GR, i.e, $f(R)=R$. Red circles indicate the threshold value to the unstable branches ($d^2f/dR^2<0$).}
\label{fRphi24e}
\end{figure}

\subsection{Quartic Potential}
One can also investigate 
\begin{equation}
V_4(\tilde\phi) = \frac{\lambda}{4} \tilde\phi^4,
\label{V4}
\end{equation}
according to which the field behaves as radiation at the final stages. It yields a similar plot to the previous one --- see Fig.~\ref{fRphi24e}, middle panel ---  except that now the second derivative $d^2f/dR^2$ is {\it not} well defined at the origin $R=0$ --- the $\tilde\phi$ mass does vanish, since $V_4''(0)=0$. 
Here, the threshold value for $\tilde\phi$ so that the unstable branch is not reached is $\beta \tilde\phi_*^{(4)} = \sqrt{3}-3 \approx -1.27$. 
Following then the same procedure as in the previous case for the new range $\beta \tilde\phi > \beta \tilde\phi_*^{(4)}$, the above potential corresponds to
\begin{equation}
f_4(R) \approx 
R + \frac{3}{4}(3\beta)^\frac{2}{3} R^\frac{4}{3} + \frac{5}{2} \left(\frac{\beta^2}{3}\right)^\frac{2}{3} R^\frac{5}{3} + {\cal O}(R^2).
\label{f4}
\end{equation}

Here, the interaction term is
\begin{equation}
{\cal L}_{\rm int}^{(4)} = 3 \xi \lambda\beta \tilde\phi^3\tilde\psi^2
= g_4 \, \tilde\phi^3 \tilde\psi^2
\label{Lint4}
\end{equation}
where $g_4 \equiv 3\xi \lambda \beta$.

\subsection{Exponential Potential}

Motivated by the functional form of Eq.~(\ref{map}), one can also investigate
\begin{equation}
V_{\rm exp}(\tilde\phi) = -\alpha (\beta \tilde\phi-a)^n {\rm e}^{-\beta \tilde\phi} + \tilde\Lambda.
\label{Vexp}
\end{equation}
The extra parameters $a$ and $\tilde\Lambda$ are chosen so that the minimum $\tilde\phi_o=0$ of the potential is at zero, i.e, $V_{\rm exp}(\tilde\phi_o)=0$ and that $df/dR|_{\tilde\phi_o}=1$, i.e, we recover GR at the end. Those constraints yield $\tilde\Lambda= \alpha n^n$ and $a=-n$. The global normalization coefficient $\alpha$ is a constant with dimensions of energy, i.e, $L^{-4}$. 

We have chosen $n=1$, that features a large plateau for large and positive $\tilde\phi$ (where the slow-roll conditions are satisfied) and a stable minimum at $\tilde\phi=0$.  Indeed, the shape of the potential given by Eq.~(\ref{Vexp}) is close to Starobinsky's 
$V_S(\tilde\phi)\equiv \frac{3}{4}M^2 [1 - \exp(-\beta\tilde\phi)]^2$, corresponding to $f(R) = R + \alpha R^2$, but with a different functional dependence on $\tilde\phi$. For an initial $\beta \tilde\phi \approx 5$, the slow-roll conditions are satisfied and the number of efolds is $\sim 60$, as expected.

For $n=1$, the corresponding function $f(R)$ is
\begin{align}
f_{\rm exp1}(R) &=
- \sqrt{\frac{3}{2}} \bigg\{\zeta + W_{-1}[-2\exp(-\zeta)]\bigg\}
\label{fexp1}\\
{\rm where~}
\zeta &\equiv \frac{R}{2\alpha} + 2
\end{align}
where $W_{-1}(\zeta)$ a non-principal branch of the Product-Log function. Notice that, although $\tilde\Lambda\neq0$ in Eq.~(\ref{Vexp}), there is no cosmological  constant in the JF, since $f(R=0)=0$.
The series expansion around $R=0$ yields
\begin{equation}
f_{\exp1}(R) \approx R + \frac{1}{2} \frac{1}{2\alpha} R^2 - \frac{1}{24\alpha^2} R^3 + {\cal O} (R^4).
\end{equation}
As before, this is also close to Starobinsky's model, as expected, but only at very low $R\ll 6\alpha$. Possible deviations between those models will be the subject of a future work.

The interaction term is similar to Eq.~(\ref{Lint2}), with the quadratic potential:
\begin{equation}
{\cal L}_{\rm int}^{\rm (exp)} = 3 \xi \alpha \beta^2 {\rm e}^{\beta \tilde\phi} (2 {\rm e}^{\beta \tilde\phi} - 2 - \beta\tilde\phi) \tilde\psi^2
\approx g_{\rm exp} \tilde\phi \tilde \psi^2
\label{Lintexp}
\end{equation}
where $g_{\rm exp} \equiv 3 \xi\alpha\beta^3$. 

\begin{figure}
\includegraphics[width=0.5\textwidth]{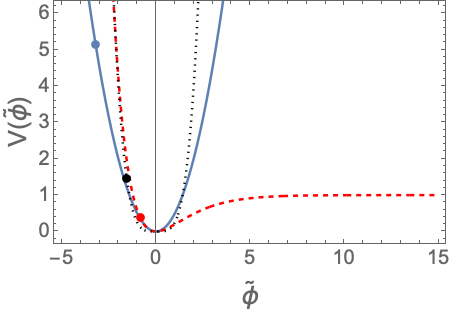}
\caption{Potentials in the EF, given by {\bf (solid blue)} Eq.~(\ref{V2}) with $\tilde m_\phi^2=1$, {\bf (dotted black)} Eq.~(\ref{V4}) with $\lambda=1$ and {\bf (dashed red)} Eq.~(\ref{Vexp}), with $n=\alpha=1$. The solid dots indicate the respective threshold values for the unstable branch of the corresponding $f(R)$ expressions.}
\label{Ves}
\end{figure}

In all panels of Fig.~\ref{fRphi24e}, one can also see the unstable branches, where $f''\equiv d^2f/dR^2<0$. The red dots (also shown in Fig.~\ref{Ves}) indicate exactly the threshold value 
between those branches. In the following calculations, we assume that the field $\tilde\phi$ starts at a large enough positive value (so that the constraints on the number of efolds and on the energy scale are satisfied). Nevertheless, its standard evolution indicates that the threshold is not crossed\footnote{The instability in this region would generate an extra amplification in the field $\tilde\psi$.} in neither case, due to the friction term in an expanding background. 
We did neglect the field $\tilde\psi$ in the equation of motion for the field $\tilde\phi$, assuming that the former is in its ground state, before preheating, i.e, during inflation. The coupling of $\tilde\phi$ to another field would only increase even further the
energy flow away from $\tilde\phi$ and its maximum oscillation amplitude would be even smaller. In other words, the unstable branch is definitely {\it not} reached.

For all the three cases above, the equation of motion for $\tilde\psi$ is written as
\begin{align}
\ddot\Psi_k - \bigg(\frac{1}{2}\beta\dot{\tilde\phi}\bigg) \dot \Psi_k + 
\bigg[
&\bigg(-\frac{9}{4}H^2 - \frac{3}{2}\dot H + \frac{k^2}{a^2} + \tilde m_\psi^2\bigg) + \nonumber \\
&- \frac{3}{4} \beta \dot{\tilde\phi} H - \frac{1}{4} \beta^2 \dot{\tilde\phi}^2 + \nonumber \\ &
 + \frac{1}{2} \beta \ddot{\tilde\phi}+ \Delta \bigg]\Psi_k=0
\label{Psik}
\end{align}
where $\Psi_k \equiv a \tilde \psi_k$, $\tilde\psi_k$ is the Fourier component of $\tilde\psi$, $H \equiv \dot a/a$ is the Hubble parameter and 
\begin{equation}
\Delta \equiv \left\{
\begin{array}{l}
2 g_2 \tilde \phi,\\
2 g_4 \tilde \phi^3 ,\\
2 g_{\rm exp} \tilde \phi,
\end{array}
\right.
\end{equation}
corresponding to the choices (\ref{V2}), (\ref{V4}) and (\ref{Vexp}), respectively.}

In the absence of the terms in $\tilde\phi$, one recovers from Eq.~(\ref{Psik}) the standard evolution equation for the modes of a scalar field $\Psi_k$ in an expanding background. Nevertheless, here the field $\tilde\phi$ exists and it does oscillate around the bottom of its potential, as discussed above. We assume that the PR happens almost instantaneously (as usual), and then neglect the expansion of the universe ($\dot H=0$, $a=1$). Therefore, the expression (\ref{Psik}) can be written as the Mathieu equation (\ref{mathieu}).

\begin{figure}
\centering
\includegraphics[width=0.45\textwidth]{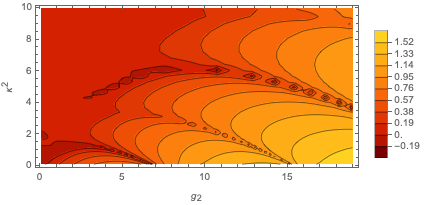}
\includegraphics[width=0.45\textwidth]{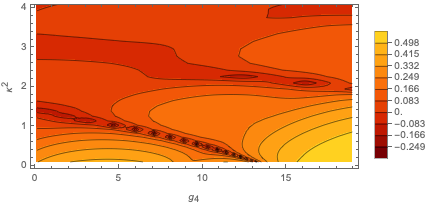}
\includegraphics[width=0.45\textwidth]{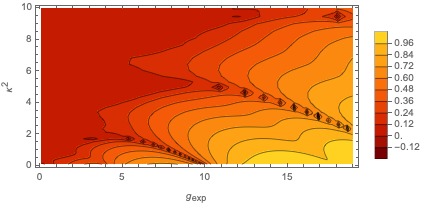}
\caption{Resonance bands for $g_2$ (upper panel), $g_4$ (middle panel) and $g_{\rm exp}$ 
 (lower panel). Legends indicate the respective range of Floquet exponents, according to the colors. In all panels, we set the initial amplitude of $\beta\tilde\phi_o=0.5$ and $\tilde m_\psi^2=1$. Note the smaller vertical range in the middle panel.}
\label{bands}
\end{figure}

Such long-term behavior of the solutions of the Mathieu Equation (\ref{mathieu})  can be depicted by the value of the so-called Floquet exponent $\mu$, estimated by
\begin{equation}
\mu = \lim_{z\to \infty} \frac{1}{2z}\ln\bigg[ \frac{A}{2}  
\bigg( |x|^2 + \frac{|dx/dz|^2}{A^2} \bigg)
\bigg],
\end{equation}
from which we are able to plot the parametric resonance bands for the exact (numerical) solutions of Eq.~(\ref{Psik}) in Fig.~\ref{bands}. 
Positive values of $\mu$ indicate exponential growth of the corresponding mode $k$.
The band structure, present in all plots, indicate that only the modes in the so-called instability bands are exponentially amplified. 
For a given model (i.e, fixed parameters $\tilde m_\psi$, $g$, etc.) each physical mode, being stretched during the expansion of the universe, will cross the bands along a vertical line, from top to bottom, going through phases of different amplification rates. The final outcome is a mix of both amplified and damped (by the universe expansion) modes, that must still be thermalized before the standard radiation-dominated universe in GR.

Meanwhile, in the JF, the Ricci scalar, which also oscillates around zero, contributes to the effective mass of the $\psi$ field through the very same mechanism of PR. Additionally, depending on the value of the $\psi$-field's bare mass $m_\psi$ --- see Eq.~(\ref{Vpsi}) --- its effective squared mass can become negative, which would then produce an extra boost to the parametric resonance alone\footnote{We recall the reader that a negative-mass-squared potential corresponds to an inverted harmonic oscillator and, therefore, to Real exponential solutions.}. Such mechanism is known as ``vacuum awakening" \cite{matsas}, triggered by $R<0$ when it is conformally coupled to a scalar field, which then grows exponentially.

Note, however, that we do require the existence of the conformal-like coupling $\sim \xi R \psi^2$ in the JF form the very beginning --- see Eq.~(\ref{Lfull}). Besides, 
 our conformal transformation (\ref{conformal}) does not include the extra scalar field $\psi$ (as opposed to Ref.~\cite{Tsujikawa:1999iv}), which means that more fields could be further included, following exact the same approach we present here, and still  one unique conformal transformation will suffice. 
 In Ref.~\cite{Faraoni:1998qx}, the authors did mention both $f(R)$ theories and conformal coupling, but no calculation of parametric resonance was made.

\section{conclusions}

In this paper we analyzed $f(R)$ theories starting from the EF (as opposed to the vast majority of papers in the current literature). Such approach supports the investigation of non-trivial modifications to GR in the JF, since they correspond to reasonable and well-motivated potentials $V(\tilde\phi)$ in the EF. 

We have made only two (reasonable) assumptions:  First, we adopt a PR-like coupling between the inflaton field $\tilde\phi$ and the secondary scalar field $\tilde\psi$. Secondly, we investigate three simple expressions for the potential $V(\tilde\phi)$ and, finally, we require from them the recovery of GR in the JF in vacuum ($f'(R\to 0)=1$). 

Although the quadratic and the quartic potentials are not strong candidates for inflationary potentials (they do not yield the right spectral index $n_s$ and tensor-to-scalar ratio $r$), they may still be useful as approximations to viable potentials close to their minimum (at $\tilde \phi=0$). The exponential potential, on the other hand, is indeed a strong candidate and its outcomes will be thoroughly investigated in a forthcoming paper.

Nevertheless, all of them present a preheating-like mechanism that leads to a hot dense universe ruled by GR in the JF. They also present an important piece of information: the Ricci scalar $R$ does oscillate around zero so it also contributes as an oscillating mass for the original field $\psi$ in the JF. 

Further outcomes from such mechanism, such as creation of primordial black holes and primordial gravitational waves as well as the evolution in the non-linear preheating phase --- thermalization and its possible signatures --- will be the subject of the next paper. 

We will further explore the similarities between the conformal transformation (\ref{conformal}) and the local gauge mechanism (and the corresponding covariant derivatives) in a future work.

SEJ thanks FAPERJ for the financial support. RA acknowledges the hospitality of Indian Institute of Science, Bengaluru, India and the financial support by National Post-Doctoral
Fellowship by SERB, Government of India (PDF/2021/004792) during the initial stages of this work. KHS thanks STDF, under Grant No. 33495, for the financial support.

	\bibliographystyle{utphys}
	\bibliography{main}
 \end{document}